\documentstyle[aps,epsf,prb]{revtex}
\begin{document}
\draft

 \twocolumn[\hsize\textwidth\columnwidth\hsize\csname@twocolumnfalse\endcsname

\title{ Suppression of ferromagnetic ordering in doped manganites:
  Effects of the  superexchange interaction }

\author{Hongsuk Yi}
\address{ Center for CMR Materials, Korea Research Institute of
  Standards and Science,\\ Yusong, P.O. Box 102, Taejon 305-600,
  Korea} 

\author{Jaejun Yu}
\address{Department of Physics and Center for Strongly Correlated Materials
  Research,\\ Seoul National University, Seoul 151-742, Korea}

\author{Sung-Ik Lee}
\address{ National Creative Research Initiative Center for
  Superconductivity, \\Pohang University of Science and Technology,
  Pohang 790-784, Korea}

\date{today}

\maketitle

\begin{abstract} 

  {}From a Monte Carlo study of the ferromagnetic Kondo lattice model
  for doped manganites, including the antiferromagnetic superexchange
  interaction ($J_{AF}$), we found that the ferromagnetic ordering was
  suppressed as $J_{AF}$ increased. The ferromagnetic transition
  temperature $T_c$, as obtained from a mean field fit to the
  calculated susceptibilities, was found to decrease monotonically
  with increasing $J_{AF}$.  Further, the suppression in $T_c$ scales
  with the bandwidth narrowing induced by the antiferromagnetic
  frustration originating from $J_{AF}$.  From these results, we
  propose that the change in the superexchange interaction strength
  between the $t_{2g}$ electrons of the Mn ions is one of the
  mechanisms responsible for the suppression in $T_c$ observed in
  manganites of the type (La$_{0.7-y}$Pr$_{y}$)Ca$_{0.3}$MnO$_3$.
\end{abstract}

 ]

\narrowtext

\section{Introduction}

Since the discovery of colossal magnetoresistance (CMR) in the doped
manganites, $Ln_{1-x}A_{x}$MnO$_3$ ($Ln$=trivalent lanthanide,
$A$=divalent alkaline), great efforts have been made to understand the
metal-insulator and/or ferro-to-antiferromagnetic transitions driven
by the substitution of cations with different sizes, such as
(La$_{0.7-y}$Pr$_{y}$)Ca$_{0.3}$MnO$_3$.\cite{hwang-prl,radaelli,garcia-prb}
For the La$_{0.7}$Ca$_{0.3}$MnO$_3$ system, i.e., ${y=0}$, which
corresponds to a hole doping of $x\sim 1/3$, a sharp resistivity peak
is observed near the transition temperature $T_c \approx 260$ K.
While the compound is insulating and paramagnetic above $T_c$, it
becomes a metallic ferromagnet below $T_c$.  The double exchange (DE)
model\cite{dex} has been widely used to describe the nature of the
metallic and the ferromagnetic (FM) states in a small range of doping
concentrations near $x \sim 1/3$.  However, it was claimed that other
effects such as the Jahn-Teller (JT)
distortion,\cite{millis-prl,rodziguez} the orbital
degeneracy,\cite{ishihara,koshibae} and the antiferromagnetic (AF)
superexchange interaction ($J_{AF}$)\cite{yunoki-prb,yi_yu,horsch}
should be included for a more accurate description over a wide range
of $x$ values $(0 \alt x \alt 1)$.

Recent experiments\cite{hwang-prl,radaelli,garcia-prb} on
La$_{2/3}$A$_{1/3}$MnO$_3$ showed that the substitution of trivalent
ions such as Y, Pr, and Dy at the La$^{3+}$ sites having a smaller
average radius, $\langle R_o \rangle$, for the lanthanide led to the
modification of the Mn-O-Mn bond angle ($\Theta$) and the Mn-O bond
length. Those experiments also revealed that these effects caused a
decrease in $T_c$ and a reduction in the magnetization below $T_c$, as
well as an increase in the CMR effects near $T_c$.  In general, the
narrowing of the electronic bandwidth is considered to be the origin
of the $T_c$ suppression effects.  In this regard, Millis {\it et
al.}\cite{millis-prl} and Rodriguez-Martinez and
Attfield\cite{rodziguez} suggested that the electron-phonon coupling
arising from the dynamic JT distortion due to the varying $\langle R_o
\rangle$ could induce the bandwidth narrowing and suppress $T_c$
rapidly.  However, it was suggested that mechanisms other than the
electron-phonon coupling could also explain the same
behavior.\cite{hwang-prl,radaelli,garcia-prb} In particular, Radaelli
{\it et al.}\cite{radaelli} observed that the suppression of $T_c$ was
more sensitive to the bandwidth narrowing due to the change in
$\Theta$ rather than to the electron-phonon coupling.  The change in
the bond angle will not only influence the DE hopping integral
($t_\sigma^{DE}$) between $e_g$(Mn)-$2p_\sigma$(O)-$e_g$(Mn) orbitals
of a FM nature but also affect the superexchange hopping integral
($t_\pi^{AF}$) through a hybridization of the
$t_{2g}$(Mn)-$2p_\pi$(O)-$t_{2g}$(Mn) orbitals of an AF
nature.\cite{garcia-prb} Thus, it is expected that both the double
exchange and the superexchange terms will compete with each other and
play a significant role when structural distortions occur.  Therefore,
it would be of great interest to study the role of $J_{AF}$ in the
doped perovskite Mn-oxides.

In this paper, we present results of our Monte Carlo calculations for
the FM Kondo lattice model which was modified by introducing the
$J_{AF}$. From the calculations of the spin correlations, we found
that the ordered FM state changes into an incommensurate (IC) state
due to the magnetic fluctuations induced by the increase of $J_{AF}$.
It was also shown from the obtained susceptibilities that the
mean-field $T_c$ monotonically decreased with $J_{AF}$.  From the
density-of-states calculations, we were able to determine that the
bandwidth narrowing induced by the AF frustration originating from
$J_{AF}$ scaled with the reduction of $T_c$. This behavior is
consistent with that seen during recent experimental observations of
the La$_{2/3-y}$Pr$_{y}$Ca$_{1/3}$MnO$_3$ system.

\section{Model Hamiltonian and Calculation Method}

The model Hamiltonian studied in this paper can be written as a sum of
three terms:
\begin{eqnarray}
\label{eq:ham}
H & = & -\sum_{\langle ij \rangle, \sigma }^{L} \Bigl( t_{ij} c_{i
  \sigma}^+ c_{j \sigma}+ {\rm H.c.} \Bigr)  - J_H \sum_{i,ab}^{L}
{\vec{S}}_i \cdot {\vec{\sigma}}_{ab} c_{ia}^+ c_{i b} \nonumber \\
  &  & + J_{AF} \sum_{i,j}^L {\vec{S}}_i \cdot {\vec S_{j}},
\end{eqnarray}
where the $\vec{S}_i$ represents the localized spin of the $t_{2g}$
electrons at the site ${\vec R_i}$, the $c_{i\sigma}$ is a destruction
operator for one species of $e_g$ fermions. The $t_{ij}$ is the
nearest-neighbor hopping amplitude for the $e_g$ electrons, and $J_H
>0$ is the FM Hund's rule coupling between the $e_g$ conduction
electrons and the $t_{2g}$ localized electrons.  The last term
represents the AF superexchange interaction between the $t_{2g}$
electrons of neighboring Mn sites.  In our calculation, we treat the
localized $t_{2g}$ spin $\vec{S_i}$ as a classical spin;
$\vec{S_i}=3/2(\sin\theta_i\cos\phi_i \hat{x}+
\sin\theta_i\sin\phi_i\hat{y} + \cos\theta_i \hat{z})$.  Within this
approximation, the trace over the $e_g$ electrons in the partition
function can be carried out exactly since Eq.~(1) is quadratic in the
fermionic field. The integration over the localized spins $\{
\theta_{\bf i}, \phi_{\bf i} \}$ is performed using a standard
Metropolis algorithm.\cite{yunoki-prb,yi_yu} In our calculation,
unless stated otherwise, we take $t=1$ as the unit of energy,
$J_H=8t$, the temperature $T=0.01t$, and $L=6 \times 6$ with a
periodic boundary condition (PBC).  Because the $J_{AF}$ is
estimated\cite{ishihara,horsch} to be on the order of $J_{AF} \alt
0.01$ eV while typical values for $t$ and $J_H$ are $ \sim 0.1$ eV and
1 eV, respectively.  The range of values for $J_H$ considered in this
work looks reasonable in comparison with those seen in experiments.

\section{Results}

\subsection{Effects of $J_{AF}$ in spin-spin correlation}

{}For a given chemical potential $\mu=-7.4$, the electron density
$\langle n \rangle$ $(= 1-x)$ was determined as a result of
simulations. In Fig.~1(a), $\langle n \rangle$ is plotted as a
function of $J_{AF}$.  While $\langle n \rangle$ stays almost constant
for the range of $J_{AF} \alt 0.11$, an abrupt jump is observed in
$\langle n \rangle$ near $J_{AF} \approx 0.115$.  For small $J_{AF}$,
the average density is about $0.65$, approximately $x \sim 1/3$, which
is almost independent of $J_{AF}$.  In this work, we focus on the
system with $x\approx 1/3$ because the FM metallic phase with large
CMR effects is observed in La$_{1-x}A_{x}$MnO$_3$ with $x\approx 1/3$
and because dramatic FM-AF and metal-insulator transitions occur when
La ions are replaced by Pr.  With increasing $J_{AF}$, the system runs
into a phase separation (PS) region where both AF and IC phases
coexist.  It should be noted that the phase separation induced by
$J_{AF}$ is similar to the discontinuity observed in the density as a
function of $\mu$.\cite{yunoki-prb,yi_yu}

\begin{figure}
\centering
\leavevmode
\epsfxsize=7.2cm\epsfbox{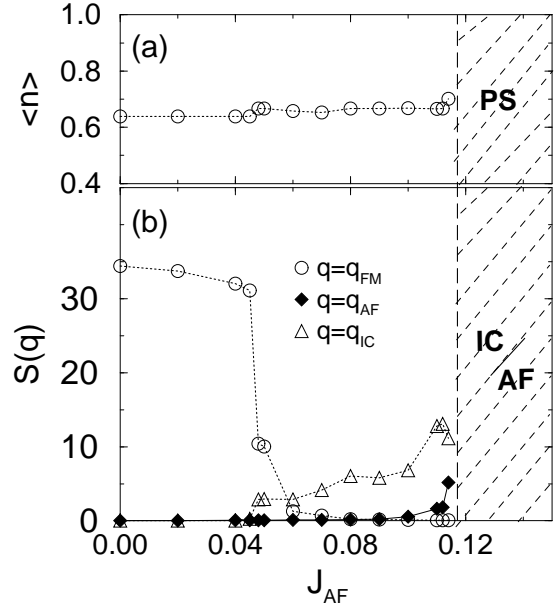}
\caption{
  (a) The carrier density vs the antiferromagnetic superexchange
  interaction $J_{AF}$ on a 6$\times$6 lattice with $J_H=8t$,
  $T=t/100$, and $\mu=-7.4$.  The notation PS means phase separation.
  (b) Peak intensities $S(q)$ for $q_{\rm FM}=(0,0)$,
  $q_{\rm IC}=(\alpha\pi,\pi)$ $(0<\alpha<1)$ and $q_{\rm AF}=(\pi,\pi)$ as a
  function of doping $J_{AF}$.}
\label{fig1} 
\end{figure}

The spin structure factor $S({\bf q})$ given by the Fourier transform
of the spin-spin correlation function, $ S({\bf q}) \equiv \sum_{\bf
i,l} e^{i{\bf q} \cdot {\bf l}} \langle {{{\vec S}_{\bf i}}\cdot{{\vec
S}_{\bf i+l}} }\rangle $, was also evaluated as a function of $J_{AF}$
and is plotted in Fig.~1(b).  The different ${\bf q}$ values of
$(0,0)$, $(\alpha\pi,\pi)$ with $(0<\alpha <1)$, and $(\pi,\pi)$
correspond to the FM, IC and AF phases, respectively.  The FM
fluctuation is dominant for $J_{AF} < 0.045$ and is drastically
reduced for $J_{AF} \agt 0.05$, where the IC component starts to
develop.  Strong IC spin correlations become significantly pronounced
with a further increase of $J_{AF}$. The presence of IC state arising
from the frustration of AF insulating state has been pointed out to
play an important role in the transport properties in
manganites.\cite{inoue} Also, near $J_{AF} \approx 0.115$, a small
region of $J_{AF}$ exists where the IC and the AF fluctuations are
competing. Due to the PS effect mentioned above, the two phases should
segregate, leading to the IC and the AF phases.  However, to
understand the fine details of the magnetic properties, the
one-orbital model is not sufficient due to the highly anisotropic
nature of the orbital degree of freedom in undoped
manganites. Certainly a two orbital model is needed to describe the
non-trivial interplay between the spin and orbital ordering in
insulating phase of LaMnO$_3$.\cite{koshibae}
\begin{figure}
\centering
\leavevmode
\epsfxsize=7.2cm\epsfbox{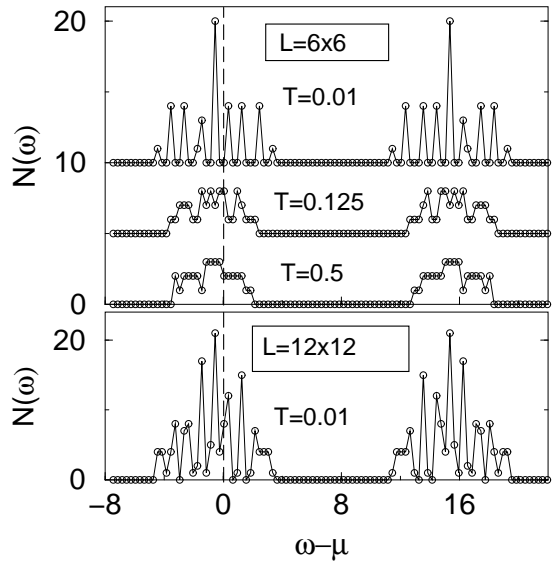}
\caption{
  Density-of-states $N(\omega)$ as a function of $\epsilon-\mu$ at
  three different temperatures with $J_{AF}=0.0$ at $x \approx 1/3$
  with a  PBC. In the lower panel, the result for a larger  system
  size of $L=12  \times 12$ is shown. }
\label{fig2} 
\end{figure}

\subsection{Bandwidth narrowing}

As shown in Fig.~2, we calculated the density-of-states $N(\omega)$,
from the eigenvalues of the Hamiltonian in Eq.~(1) for many
configurations at three different temperatures and system sizes with
$J_{AF}=0.0$. The broken line at $\omega-\mu=0$ indicates the Fermi
energy $E_F$.\cite{omega} It can be seen that the bands are split into
two major parts with the separation between the lower and the upper
bands corresponding to the energy splitting $2J_H$ by the Hund
coupling.  In the low-temperature limit ($T=0.01$), the spiky shape of
each band is due to the finite-size effect, as shown by the
calculations for two system sizes, $L=6\times 6$ in the upper panel
and $12\times 12$ in the lower panel. As the temperature increases,
this shape is smeared out, and at the same time the width of each band
is reduced due to the disorder of the local $t_{2g}$ spins.  The
system is FM metallic since the density-of-states at $E_F$ remains
finite.

\begin{figure}
\centering
\leavevmode
\epsfxsize=7.2cm\epsfbox{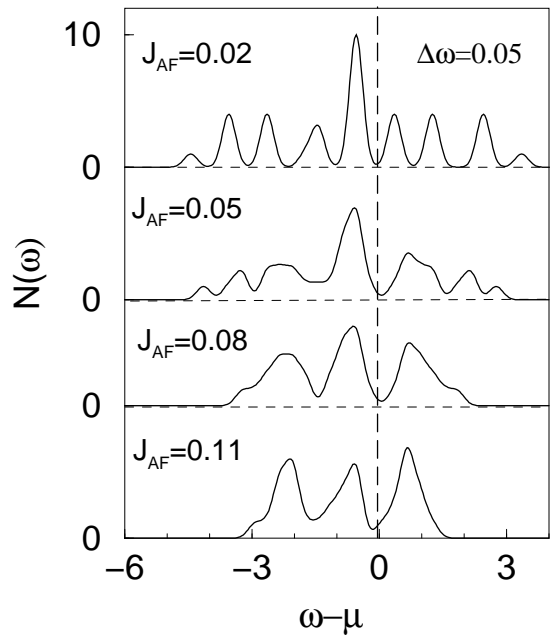}
\caption{
  Density-of-states as a function of $\epsilon-\mu$ for $J_{AF}=0.02,
  0.05, 0.08$ and $0.11$.}
\label{fig3} 
\end{figure}

The $N(\omega)$ of the lower band is shown for various values of
$J_{AF}$ in Fig.~3. These curves are smoothed out with a Gaussian
broadening of $\Delta \omega = 0.05$.  This broad spectrum of
$N(\omega)$ is also observed for $0 \alt J_{AF} \alt 0.04$, which
corresponds to the FM region in Fig.~1(b).  With further increase in
$J_{AF}$, we find that a significant difference exists in the nature
of each band between the FM and the IC states.  For $0.05 \alt J_{AF}
\alt 0.11$, together with a strong band narrowing, the wide
distribution of eigenvalues merge into three peaks while the $E_F$
lies at the dip of $N(\omega)$.

In Fig.~4, the bandwidth determined from the density-of-states is
plotted as a function of $J_{AF}$. We define the bandwidth of the
lower band as $\Delta W \equiv \langle \epsilon_{max} - \epsilon_{min}
\rangle$ where $\epsilon_{max}$ and $\epsilon_{min}$ are the maximum
and the minimum energies of each band in Fig.~2.  Surprisingly, the
bandwidth has a unique dependence on $J_{AF}$ in each magnetic region,
as discussed above. In the metallic FM region, the bandwidth is
independent of $J_{AF}$ and is equal to about 8.  This values is close
to the one from the dispersion relation in the 2-dimensional DE model.
The most dramatic change of $W$ with $J_{AF}$ is found in the metallic
IC region. $W$ monotonically decreases by half as $J_{AF}$ increases
from 0.04 to 0.11. The evolution of the bandwidth depicted in Fig.~4
is of great importance since the suppression of $T_c$ in the doped
manganites is closely related to the bandwidth narrowing.  It implies
that the AF frustration arising from $J_{AF}$ can be an another
candidate for the band narrowing mechanism in the doped manganite
systems besides the JT-polaron effects\cite{millis-prl} and the random
potentials due to the disorder of the cations.\cite{rodziguez}

\begin{figure}
\centering
\leavevmode
\epsfxsize=7.2cm\epsfbox{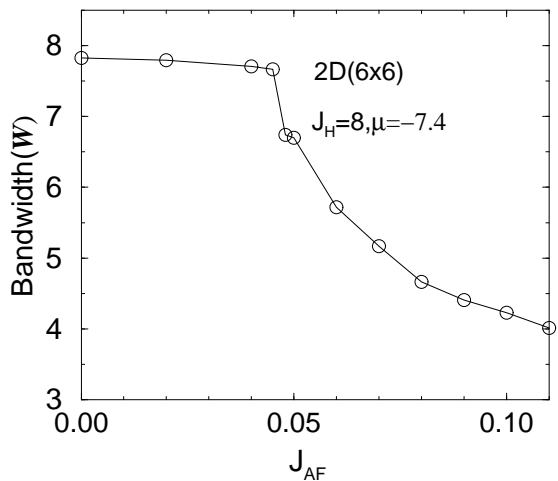}
\caption{
  Bandwidth as a function of $J_{AF}$ for $J_H=8$ and $T=0.01$ with a 
  PBC. }
\label{fig4} 
\end{figure}

\subsection{Suppression of ferromagnetic ordering}

The temperature-dependent magnetic susceptibility is calculated from
$\chi(T) \equiv\langle S_z^2 \rangle/k_BT$.\cite{horsch} In Fig.~5,
the inverse $\chi(T)$ is plotted vs temperature for various values of
$J_{AF}$; a Curie-Weiss behavior is seen for $\chi(T)$ at high
temperatures.  The mean-field transition temperature $T_M$ is
estimated by extrapolating the high-temperature data to the
low-temperature limit.  It is found that $T_M$ decreases from positive
to negative values, indicating a change of the interaction nature from
FM to AF, with increasing $J_{AF}$. The change of $T_M$ with $J_{AF}$
implies that the magnetic interaction via the DE hopping mechanism is
weakened by increasing the $J_{AF}$ interaction.

\begin{figure}
  \centering \leavevmode 
  \epsfxsize=7.2cm\epsfbox{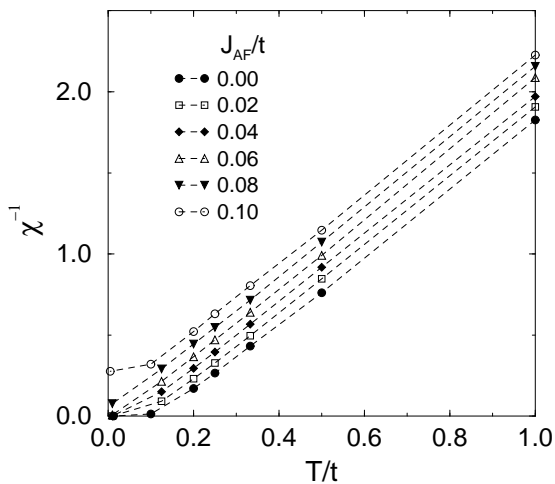}
\caption{
  Inverse susceptibility vs temperature for various values of 
  $J_{AF}$ with $J_H=8t$, $T=t/100$, and $\mu=-7.4$.  }
\label{fig5} 
\end{figure}

Based on the estimated results of $T_M$ in Fig.~5, the mean-field
temperature vs $J_{AF}$ phase diagram is shown in Fig.~6. $T_M$ in the
FM region is determined for two boundary conditions; a PBC (circles)
and open boundary condition (OBC) (squares). The AF region (triangles)
is determined using only a PBC. The ferromagnetic $T_M$ decreases
linearly within the error limit, with increasing $J_{AF}$ and reaches
zero at $J_{AF} \approx 0.08$ above which the AF $T_M$ start to
develop. The inset shows an experimentally proposed phase
diagram\cite{hwang-prl} for La$_{0.7-y}$Pr$_{y}$Ca$_{0.3}$MnO$_3$ as a
function of Pr doping. A decrease in $T_c$ when Pr doping increases
implies a weaker DE hopping overlapping and a narrower bandwidth. The
two phase diagrams shown in Fig.~6 exhibit a similarity between the Pr
doping level and $J_{AF}$.  As pointed out earlier, the substitution
of Pr at La sites causes a smaller $\langle R_o \rangle $ and a
smaller bond angle $\Theta$, and the change in $\Theta$ modifies both
the $t_\sigma^{DE}$ due to the nature of $dp\sigma$ hybridization and
the $t_\pi^{AF}$ due to $dp\pi$-type hybridization. However, while $t$
is more sensitive to a change of bond angle $\Theta$ due to the nature
of $dp\sigma$ hybridization, the superexchange term $J_{AF}$, which
relies on a $dp\pi$-type hybridization between
$t_{2g}$(Mn)-$p_{\pi}$(O)-$t_{2g}$(Mn), is relatively insensitive to a
change in the bond angle. Thus, an increase in $J_{AF}$, relative to
the DE hopping integral $t$, is expected when $\Theta$ decreases.
This behavior is consistent with the experimental
evidence\cite{garcia-prb} seen in the reduction of the magnetization
with $\langle R_o \rangle $.

\begin{figure}
\centering
\leavevmode
\epsfxsize=7.2cm\epsfbox{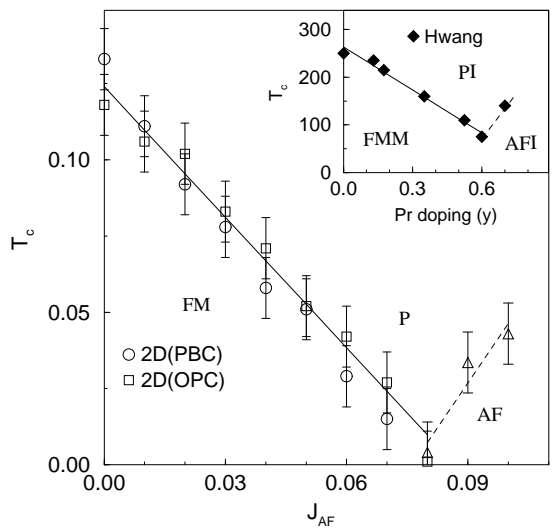}
\caption{
  Phase diagram of the 2D FM Kondo lattice model with $J_{H}=8t$,  as
  obtained by measuring the inverse susceptibilities at $x \sim 1/3$.
  The circles (squares) and triangles represent the phase boundaries
  between the paramagnetic (P) and the FM phases with a PBC (OBC) and
  between the P and the AF phases with a PBC respectively.  The inset
  shows the results for   La$_{0.7-y}$Pr$_{y}$Ca$_{0.3}$MnO$_3$ from Ref.~1.
  The solid and the dash lines  are guides for the eyes.}
\label{fig6} 
\end{figure}

Indeed, when the bandwidth in Fig.~4 is compared with $T_c$ in Fig.~6,
the suppression in $T_c$ for La$_{0.7-y}$Pr$_{y}$Ca$_{0.3}$MnO$_3$
correlates well, at least qualitatively, with the narrowing of the
bandwidth.  Further, the absence of internal spin structure, i.e., IC
ordering, in Fig.~6 for $J_{AF} \alt 0.08$ is due to the mean-field
approximation in determining $T_M$ for $\chi$.  Thus, the bandwidth
narrowing in the metallic IC state is thought to be responsible for
the observed $T_c$ reduction. Consequently, in addition to the
electron-phonon interactions arising from Jahn-Tell distortion, the
magnetic frustration arising from the enhancement of $J_{AF}$ should
be considered as a cause of the bandwidth narrowing leading to the
suppression in $T_c$.  Here, it is noted that the effect of orbital
degeneracy is not taken into account in this work.  Since, however, many
optimally doped metallic systems near $x\approx 0.3$ are pseudo-cubic in
structure, the conduction electron hopping can be considered to be
isotropic.   On the other hand, the non-trivial interplay between spin and
orbital degree of freedom in connection with orbital degeneracy has
been pointed out to play an important role in the spin and orbital
ordering in the undoped insulating phase of LaMnO$_3$.\cite{ishihara}

\section{Conclusion}

In conclusion, it is suggested that the superexchange interaction in
the modified FM Kondo lattice model is one of the mechanisms
responsible for the observed $T_c$ reduction in
La$_{0.7-y}$Pr$_{y}$Ca$_{0.3}$MnO$_3$. With a change in $J_{AF}$,
magnetic frustration is induced and leads to a more complicated
magnetic phase diagram. The bandwidth narrowing occurs for $ 0.05 \alt
J_{AF} \alt 0.11$ in the metallic IC state and is thought to be
responsible for the $T_c$ reduction in
La$_{0.7-y}$Pr$_{y}$Ca$_{0.3}$MnO$_3$. With our model, we found that
bandwidth narrowing, accompanied by $T_c$ reduction, can be induced by
$J_{AF}$ without electron-phonon coupling.

\acknowledgements

 This work was supported by Creative Research Initiatives of the
Korean Ministry of Science and Technology. Additional support was
provided by the Korea Science Engineering Foundation
(95-0702-03-01-3). Part of the calculation was also performed on the
JRCAT Supercomputer System, which is supported by New Energy and
Industrial Technology Development Organization (NEDO) of Japan.


\begin{references}
\bibitem{omega} The finite energy interval used for the
  density-of-states calculation is $\omega_o=0.3$.
\end{references}
\end{document}